\documentclass{iopjournal}

\usepackage{multirow}%
\usepackage{amsmath,amssymb,amsfonts, mathtools}%
\usepackage{bm}
\usepackage{amsthm}%
\usepackage{mathrsfs}%
\usepackage[title]{appendix}%
\usepackage{textcomp}%
\usepackage{manyfoot}%
\usepackage{booktabs}%
\usepackage{algorithm}%
\usepackage{algorithmicx}%
\usepackage{algpseudocode}%
\usepackage{listings}%
\usepackage{tikz}
\usepackage{subcaption}

\begin{document}

\articletype{Paper} 

\title{Topology of Shape and Data in Material Microstructures}

\author{Jeanie Schreiber$^{1,2}$\orcid{0000-0000-0000-0000}, Zachary Grey$^2$\orcid{0000-0000-0000-0000} and Adam Creuziger$^{3,4}$\orcid{0000-0001-9772-2066}}

\affil{$^1$Mathematical Sciences, George Mason University, Fairfax, USA}

\affil{$^2$ITL, National Institute of Standards \& Technology, Boulder, USA}

\affil{$^3$Materials Science and Engineering Division, National Institute of Standards \& Technology, Gaithersburg, USA}

\affil{$^4$Metallurgical and Materials Engineering, Colorado School of Mines, Golden, USA}

\email{jschrei@gmu.edu, zachary.grey@nist.gov, adam.creuziger@nist.gov}


\begin{abstract}
One of the challenges in microstructure analysis is the rigorous quantification of the shape, size, and spatial arrangement of the microstructure beyond comparison of average values.  We expound on formal principles combining Topological Data Analysis (TDA) and non-Euclidean distances between curves to motivate novel perspectives on the form and nature of pattern and shape in images. Specifically, TDA descriptors extracting persistent topological structures are combined with product submanifold learning of separable shape tensors (SST) to offer unique insights about electron backscatter diffraction (EBSD) images of material microstructures through the lens of a dual-parameter filtration. Beyond standard approaches, our methodology highlights how different choices or permutations of shape distances can lead to distinct notions of topological persistence, thereby broadening the interpretive scope of TDA. The resulting visualizations of feature extraction are designed to be both principled and explanatory, offering novel tools for modern imaging science with applications to material metrology. More broadly, this framework has strong potential to impact domains where precise quantification of topology and shape is critical for uncovering fundamental image patterns and features, and enables additional data-driven tools for microstructure analysis.
\end{abstract}

\section{Introduction}\label{sec1}

The precise measurement and quantification of shapes and the arrangement of these shapes in space are fundamental problems in image analysis, with broad applications including engineering design~\cite{Grey2017,Doronina2023}, manufacturing of materials~\cite{atindama2023restoration,bachmann2010inferential}, medical imaging~\cite{mang2019claire,durrleman2014morphometry}, sensing and functional analysis~\cite{tucker2013generative,tucker2011coherence}, and biology~\cite{hartman2023elastic,srivastava2010shape,hagwood2013testing}. In many scenarios of image analysis, scientists and engineers must determine whether two images are the same or different and, if different, also understand why they differ. While humans are often quite good at noticing visual patterns and recognizing differences, translating expert knowledge to a robust algorithm or machine learning framework has proved more difficult. Often, improved methods which include the development of holistic descriptors that enable precise comparison and measurement of patterns in imaging data are needed.

In materials science, a key application of image analysis is in the analysis of the microstructure. Most engineering materials are made up of millions of individual domains with a common crystal structure and crystal orientation, described as a material `grain' or polycrystal.  The grain size, grain shape, and arrangement of the grain shapes can provide information about the prior history of the material, as well as affecting how the material may perform in service in expected strength or chance of failure. 

Size and shape can be measured with well-established tools, but the arrangement of the grain shapes (i.e. topology) has been more difficult to quantify. In contrast to geometric shape representations and statistical ensembles, topological descriptions focus on the global connectivity of a dataset, offering an abstract view of its arrangement that is independent of precise geometry. Topological analyses capture global characteristics answering: ``How many distinct pieces (connected components) exist in the data?'' ``How many loops (cycles) does the structure form?'' ``Are there enclosed spaces (holes) within the data?'' 
Persistent homology has emerged as a powerful computational tool for extracting topological features directly from image data, with applications ranging from medical image segmentation~\cite{ peng2024phg, franccois2024train}, biological morphology quantification~\cite{li2018persistent}, histopathology analysis~\cite{qaiser2019fast}, and other complex imaging modalities~\cite{pun2018persistent, garin2019topological, turkevs2021noise}.

\begin{figure*}[ht]
    \centering
    \includegraphics[width=0.46\textwidth]{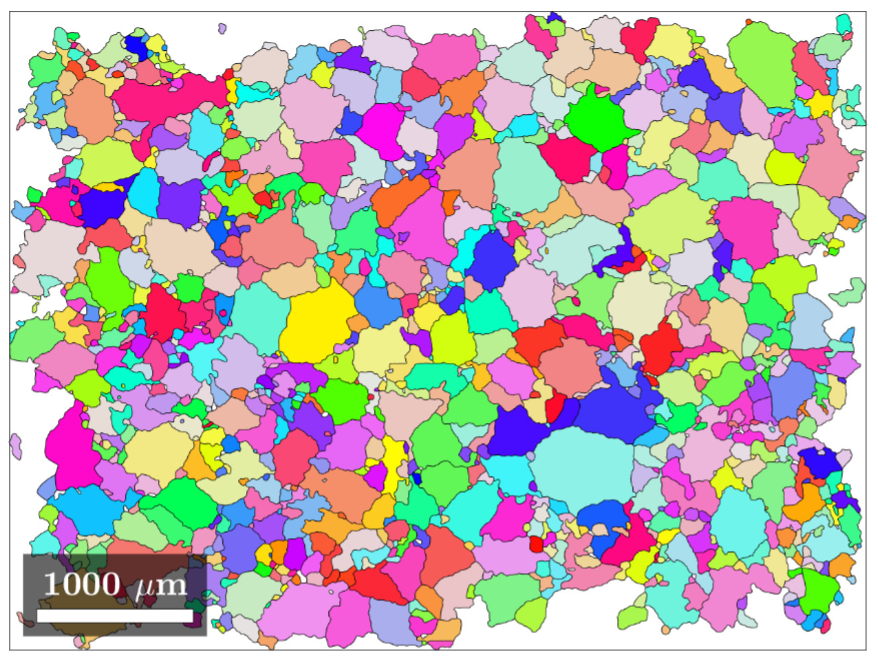}
    \hspace{4mm}
    \includegraphics[width=0.46\textwidth]{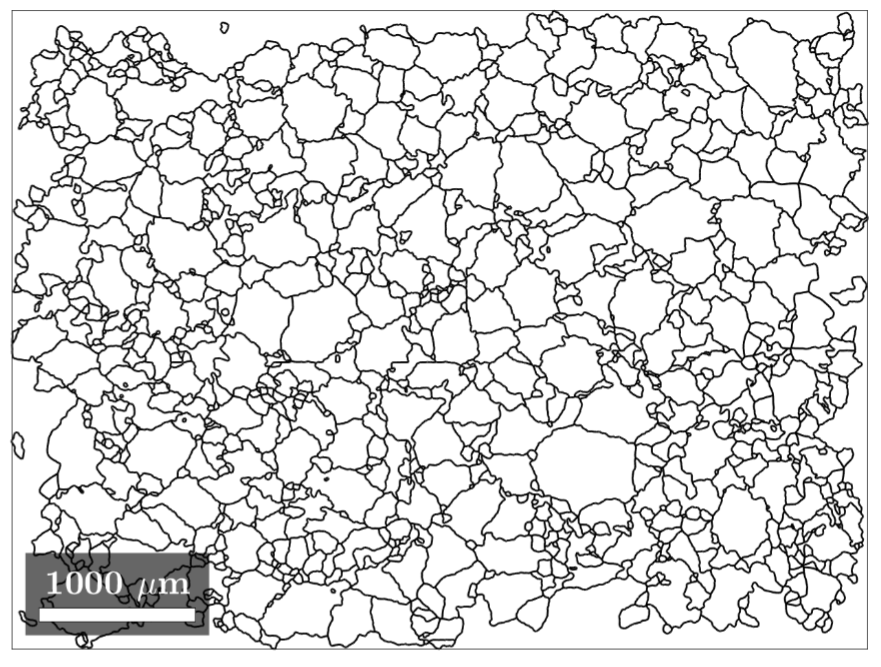}
    
    \caption{An example microstructure (PIL184  from~\cite{Fan2020}) recorded via EBSD.  The left image shows the orientation data, and the right image shows the grain boundary curves extracted from local changes in the orientation data.   
    }
    \label{fig:orientation_and_boundaries}
\end{figure*}

Electron backscatter diffraction (EBSD)~\cite{schwartz2009electron} imaging is a newer technique to image material microstructures that informs a robust demarkation of the grain boundaries by use of the orientation data~\cite{SAVILLE2021102118,stoudt2020location}. Modern computer vision technologies have also been applied to EBSD data to denoise images~\cite{atindama2023restoration} and extract~\cite{bachmann2010texture} large ensembles of planar curves from EBSD grain boundary data. An example of a segmented EBSD image from~\cite{Fan2020, fan2021using}, processed utilizing an open-source software called MTEX~\cite{bachmann2010texture, Hielscher:cg5083}, is shown in Figure~\ref{fig:orientation_and_boundaries}. In this image, regions of common crystal orientation (illustrated by the colors shown on the left of Figure~\ref{fig:orientation_and_boundaries}) are defined as grains of the microstructure. Grain boundaries are determined by orientation changes in adjacent pixels, which provide an unambiguous segmentation of the image in comparison to the more common methods in grain segmentation using grayscale value. These grains boundaries are closed planar curves, shown on the right of Figure~\ref{fig:orientation_and_boundaries}. 
While orientation data and analysis is a key feature of EBSD imaging, this work is focused solely on the character of grain boundaries and their spatial arrangement in an effort to quantify ``texture'' in imaging data, which refers to the spatial arrangement and quantitative representation of shapes forming patterns. Henceforth, the term `texture' used throughout this paper shall mean topological texture, and not the crystallographic texture often associated with EBSD orientation data.

\begin{figure*}[ht]
    \centering
    \includegraphics[width=0.24\textwidth]{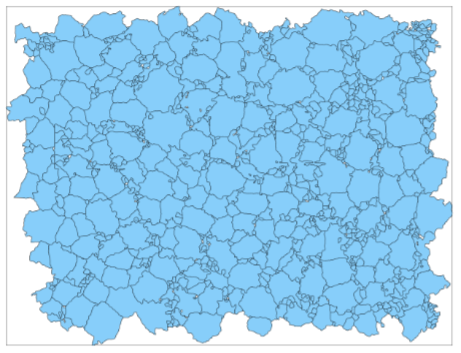}
    \includegraphics[width=0.24\textwidth]{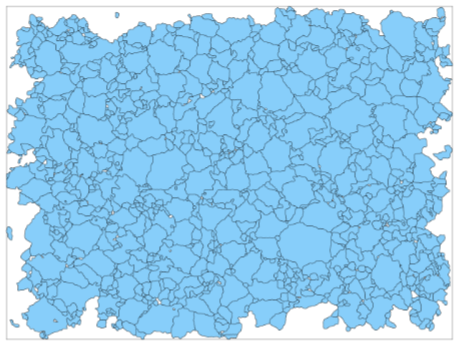}
    \includegraphics[width=0.24\textwidth]{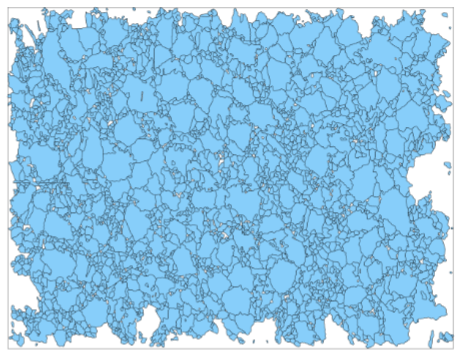}
    \includegraphics[width=0.24\textwidth]{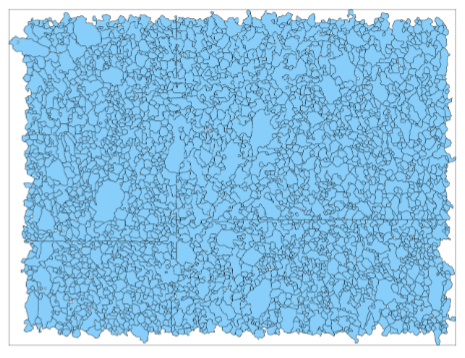}
    
    \caption{Four example microstructures of ice samples PIL254 (0.03 $\frac{mm}{mm}$), PIL184 (0.08 $\frac{mm}{mm}$), PIL185 (0.12$\frac{mm}{mm}$), and PIL255 (0.20$\frac{mm}{mm}$) from~\cite{Fan2020, fan2021using} that have solidified, deformed, and allowed to recrystalize under different conditions. Only grain boundaries (black lines) and grains included in the data set (cyan fill) are shown.  
    }
    \label{fig:eg_grains}
\end{figure*}

A practical example of the analysis challenge is shown in Figure~\ref{fig:eg_grains}.  In data from~\cite{Fan2020, fan2021using}, ice microstructures that have been solidified, deformed, and allowed to recrystalize under different conditions were imaged using EBSD. The succession of images show the resulting microstructures after being subjected to increasing axial strain ($\frac{mm}{mm}$) deformation, from PIL254 (0.03 $\frac{mm}{mm}$), PIL184 (0.08 $\frac{mm}{mm}$), PIL185 (0.12 $\frac{mm}{mm}$), and PIL255 (0.20 $\frac{mm}{mm}$)
 ~\cite{Fan2020, fan2021using}. The initial microstructure on the left is dominated by larger grains with a few clusters of small grains. As the deformation continues, numerous smaller grains form. These smaller grains are distributed in rings at boundaries of larger grains, resulting in notably different topology.  This structure of small grains surrounding large grains is akin to the necklace structure defined in ASTM E1181~\cite{ASTM_E1181}.

While the progression of shape and topological feature changes in Figure~\ref{fig:eg_grains} are apparent, they are difficult to quantify. Starting with consideration of shape, methods often lack sensitivity to higher-order spatial patterns and may be ambiguous
or inconsistent when confronted with complex morphologies (e.g., those exhibiting both spatial and geometric heterogeneity). Researchers frequently rely on qualitative assessments or conduct quantitative comparisons based on summary statistics. Mean or maximum values of a distribution (examples such as \cite{ASTM_E112, ASTM_E1382} for grain size) are often extracted as the primary values to compare. 
Methods to quantify volume content in the material of an inclusion or secondary crystal structure (\textit{i.e.} phase)~\cite{ASTM_E45, ASTM_E562, ASTM_E1245} are also widely used. Many of these hand-picked features have been linked to key properties of materials, however there is no evidence to suggest that they are comprehensive nor representative, and summary statistics that compare them can be misleading, \emph{i.e.} a phenomenon akin to Anscombe's quartet~\cite{anscombe1973graphs}.

Separable Shape Tensors (SSTs) offer a more complete characterization of grain boundary geometry by mapping individual curves onto matrix manifolds that encode both linear and nonlinear deformations~\cite{grey2023separable}. These manifold-valued representers converge to complete descriptions of shape geometry in the sense that they capture all information about the curve's intrinsic structure modulo rigid transformations. Unlike hand-picked 
descriptors that reflect implicit assumptions about relevance, SST features are derived systematically from the 
spectral decomposition of the curves themselves, ensuring that no geometric information is omitted a priori.

However, while SSTs provide a rigorous means of comparing grain shape distributions between two microstructure images~\cite{grey2024explainable}, their encodings remain completely agnostic to the broader arrangement of the grains themselves. Microstructures exhibiting similar SST shape distributions can 
have drastically different spatial organizations, which materials scientists recognize as critical but struggle 
to quantify with existing tools. We propose a novel framework that merges Topological Data Analysis (TDA) with shape analysis by combining topological persistence with shape distances---\emph{e.g.}, SST product submanifold distances. The exploration leverages biparameter persistence analysis of a spatial TDA filtration parameter conditioned on shape distances from an intrinsic, archetypal (origin) shape. A dual-parameter approach not only considers the persistence of topological features across different scales but also evaluates features' geometric similarity. In practical terms, this allows scientists to ask not only ``Are these two random ensembles of shapes from the image the same?" but also ``How are they arranged, and what global patterns emerge from their arrangement?" 

This richer analysis is particularly important in fields like materials science, where understanding more comprehensive notions of texture in a material's microstructure can be crucial for determining subsequent macroscopic properties and performance. \textit{The present discussion is an attempt to guide this measurement process with methods that combine shape analysis and computational topology.}

\subsection{Background}

Existing standards for measurement of microstructures generally focus on quantifying intuitive characteristics of shapes alone~\cite{ASTM_E112, ASTM_E1382, ASTM_E45, ASTM_E562, ASTM_E1245} and ignore the topology of the spatial arrangement of those shapes. A few select standards do address topology, by considering the largest grain \cite{ASTM_E930} or morphologies observed in duplex grain sizes \cite{ASTM_E1181}.
However, even when topology is mentioned, methods to analyze or report the spatial arrangement are limited. In ASTM E45-``Standard Test Methods for Determining the Inclusion Content of Steel''~\cite{ASTM_E45}, the `distribution' of inclusions is mentioned as a characterization goal, but no mention on how the distribution might affect the `severity number' or how the distribution shall be recorded in the expression of results is given.  In ASTM E1181-``Standard Test Methods for Characterizing Duplex Grain Sizes''~\cite{ASTM_E1181}, `topologically varying' is explicitly defined as ``varying nonrandomly, in some definable pattern; that pattern may be related to the shape of the specimen or product being examined.''  Two named examples are provided. `Banding', where there are alternating areas of different grain sizes, often with the areas and grain shapes elongated along the direction of work. `Necklacing', where ``individual coarse grains are surrounded by rings of significantly finer grains.'' Additional criteria are discussed in the `Procedure' section 8.3.2 of ASTM E1181~\cite{ASTM_E1181}, but example figures are relied on to qualitatively describe the topology and grain size distribution. There is a gap where more quantitative tools to describe these structures are needed.

In computer assisted analysis of microstructures, additional metrics of shape are possible beyond what is considered in documentary standards. A wide variety of software is available for these---\emph{e.g.}, see the list of ``Shape Parameters" from MTEX~\cite{Hielscher:cg5083}. However, a variety of distinct implementations, over-simplification of shape representations (\emph{e.g.}, fitting a circle or ellipse), and emphasis on extracting a single metric or value (such as grain size) limit comparisons of microstructures and are an opportunity to develop additional comparison methods and standards.

In~\cite{grey2024explainable}, kernel metrics over learned SST features are used to construct an explainable binary classifier for shape ensembles, capable of effectively distinguishing between microstructure images such as those in Figure~\ref{fig:eg_grains}. Two probability measures over SST features differ precisely by virtue of the metric constructed through deliberate choices of geometric invariance, and the separability of the tensor decomposition allows individual shape factors to be examined without conflating distinct geometric phenomena. However, this framework does not account for the spatial arrangement of grain shapes within a microstructure. The present work seeks to augment this analysis by quantifying the inherent topological structure of microstructures, exploring patterns in spatial arrangement conditioned on shape similarities (distances) encoded over the SST submanifold.

For broad imaging applications, TDA can reveal significant topological features that are not immediately apparent from geometric measurements alone~\cite{garin2019topological, singh2023topological}. Existing TDA methods are especially attractive for facilitating explanations since they provide detailed summaries of the global structure through the lens of persistent homology. However, the majority of current TDA applications focus on single-parameter persistence, relying on a one-dimensional filtration to index changes in topology. This restriction to a single parameter undermines their effectiveness compared to other, more expressive, data-driven methods.  

Recently, theoretical developments have extended the machinery of persistent homology from one-dimensional filtrations to multiparameter settings (often called multi-parameter persistent homology) where topological features are tracked across two or more simultaneously varying parameters~\cite{botnan2022introduction,carlsson2007theory}, including applications to three dimensional imaging~\cite{vipond2021multiparameter} and image denoising~\cite{chung2022multi}. For instance, the Mix-GENEO~\cite{he2024mix} filtration introduces a flexible framework for constructing bifiltrations on digital images by combining grayscale intensity with spatial proximity. While Mix-GENEO focuses on pixel-level intensity for general image classification, our approach adapts the multiparameter philosophy to material microstructures by pairing spatial Rips filtrations with manifold-valued SST shape distances.

The remainder of this paper is organized as follows: Section~\ref{sec2} reviews related work in topological data and shape analysis methods. Section~\ref{sec4} details our proposed framework, including the integration of SST with TDA descriptors. Section~\ref{sec5} presents experimental results and evaluations on both synthetic and real-world data sets. Finally, Section~\ref{sec6} discusses the implications of our findings and Section~\ref{sec7} future research directions.

\subsection{Motivation \& Contribution}

We introduce a novel method that augments a TDA filtration with shape distance sublevel sets from an archetypal origin to gauge biparameter persistence. This approach allows us to uncover topological patterns which persist over a continuum of similar shapes; enabling a direct application to the quantification of `necklacing' structures described in ASTM materials science standards as one illustrative case. Applied to EBSD scans of ice recrystallized under increasing axial strain (0.03 to 0.20 $\frac{mm}{mm}$), our scalar bipersistence summary $\mathcal{I}$ reveals a sharp transition in topological texture coinciding with the onset of dynamic recrystallization as the dominant grain-refinement mechanism. This trend is consistent with prior observations by Fan et al~\cite{fan2020temperature} on the evolution of ice microstructure under progressive strain, and it illustrates the kind of mechanistic insight that existing summary statistics cannot provide.

Although the present demonstration uses EBSD scans of recrystallized ice, the underlying construction depends only on grain-centroid positions and SST-derived shape distances, both of which are routinely extractable from EBSD or optical micrographs of metals, ceramics, and composites. The framework is therefore directly transferable to characterizing necklacing, banding, and duplex grain structures in structural alloys and other polycrystalline engineering materials where ASTM E1181-type qualitative assessments are currently standard practice.

Broadly, the aim is to motivate novel, holistic quantification of coherent patterns in scientific imaging. While we demonstrate an application to axial strain in ice material, our goal is not to optimize classification accuracy for a specific material property prediction task. Instead, we demonstrate that biparameter persistent homology, grounded in shape-distance filtrations, enables quantitative and interpretable mapping of diverse topological features in images, allowing for new insights into the intrinsic structure and variation of texture across scientific and industrial domains.

We remark that existing approaches leveraging artificial intelligence to accomplish similar tasks are often confounded by ambiguous \textit{latent spaces} which can be difficult to interpret and explain. In contrast, every axis and summary statistic in our construction is directly traceable to a physically interpretable quantity, so a materials scientist can identify precisely which grains and spatial scales drive an elevated signal. Thus, there is no clear comparison to make since modern machine learning often falls short of explaining which patterns are intrinsic and significant in an image.

\section{TDA and Shape Analysis}\label{sec2}
\subsection{Topological Data Analysis}

Following the developments of~\cite{edelsbrunner2022computational,carlsson2009topology,zomorodian2004computing,hatcher2002algebraic}, TDA algorithms take advantage of the group structure of homological algebra to emphasize qualitative features of data such as connectivity and holes\footnote{Throughout this paper, `hole' refers to the topological notion of $1$-dimensional void structure quantified by Betti numbers $\beta_1$. This differs from the materials science definition of `hole' as a physical defect or pore in the material.}. Rather than fitting a parametric model or computing summary statistics, TDA extracts a coordinate-free description of global connectivity. The output is a multi-scale topological summary
that requires no prior assumptions about data geometry.

In most applications, TDA algorithms are used to extract so-called \textit{persistent} topological features. Betti numbers are the most common example that have proven useful in data analysis, as well as the related Euler characteristic which has been used in shape analysis~\cite{meng2024randomness}. These descriptors give a holistic summary of the topology of data and are easily scaled to higher dimensional data.

The foundational descriptors used throughout this work are the Betti numbers. Informally, the $k$-th Betti number $\beta_k$ counts the number of independent $k$-dimensional `holes' in a topological space. In practice, two Betti numbers dominate applications to spatial data: $\beta_0$ counts the number of connected components (distinct, disconnected
clusters of points), and $\beta_1$ counts the number of one-dimensional loops or cycles of closed chains of points that encircle an empty region without being filled in. 

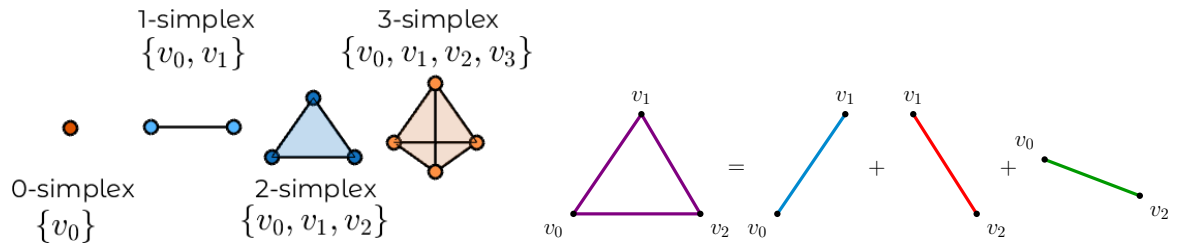
\begin{figure}
    \includegraphics[width=0.45\linewidth]{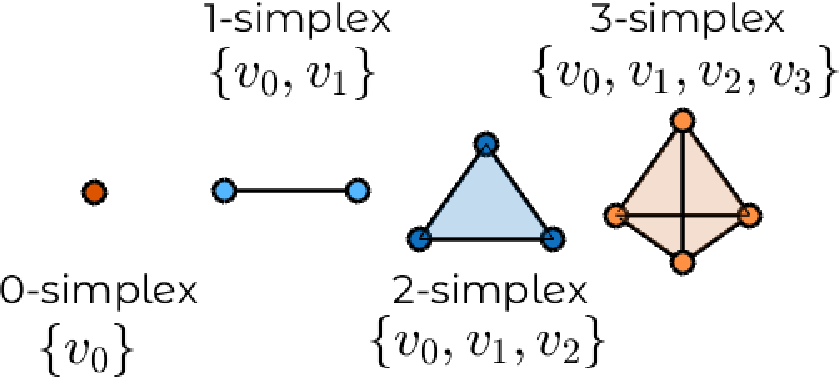}
    \resizebox{0.55\linewidth}{!}{\begin{tikzpicture}[>=stealth, line cap=round, line width=2pt,
                    every node/.style={font=\Large}]
  \coordinate (v0L) at (0,0);
  \coordinate (v1L) at (1.5,2.2);
  \coordinate (v2L) at (2.8,0);

  \coordinate (v0B) at (4.5,0);
  \coordinate (v1B) at (6,2.2);

  \coordinate (v1R) at (7.5,2.2);
  \coordinate (v2R) at (8.9,0);

  \coordinate (v0G) at (10.4,1.2);
  \coordinate (v2G) at (12.5,0.4);

  \draw[draw=violet] (v0L) -- (v1L) -- (v2L) -- cycle;
  \fill (v0L) circle (2pt);
  \fill (v1L) circle (2pt);
  \fill (v2L) circle (2pt);

  \node[below left=2pt] at (v0L) {$v_0$};
  \node[above=2pt]      at (v1L) {$v_1$};
  \node[below right=2pt]at (v2L) {$v_2$};

  \node at (3.6,1) {\Large$=$};

  \draw[draw=cyan!70!blue] (v0B) -- (v1B);
  \fill (v0B) circle (2pt);
  \fill (v1B) circle (2pt);
  \node[below left=2pt] at (v0B) {$v_0$};
  \node[above=2pt]      at (v1B) {$v_1$};

  \node at (6.7,1) {\Large$+$};

  \draw[draw=red] (v1R) -- (v2R);
  \fill (v1R) circle (2pt);
  \fill (v2R) circle (2pt);
  \node[above=2pt]      at (v1R) {$v_1$};
  \node[below right=2pt]at (v2R) {$v_2$};

  \node at (9.6,1) {\Large$+$};

  \draw[draw=green!60!black] (v0G) -- (v2G);
  \fill (v0G) circle (2pt);
  \fill (v2G) circle (2pt);
  \node[above left=2pt] at (v0G) {$v_0$};
  \node[below right=2pt]at (v2G) {$v_2$};
\end{tikzpicture}}

\vspace{3mm}
\caption{Examples of simplices of various dimension (left) and an example 1-chain (right). The 1-chain $\{v_0,v_1\}+\{v_1,v_2\}+\{v_2,v_0\}$ on the right forms a closed loop traversing three edges. The 2-simplex $\{v_0,v_1,v_2\}$ shown in blue on the left contains the same three edges as its boundary, but the interior is filled. When this triangle is present in the complex, the loop does not encircle a genuine hole since it is the boundary of a face (not a loop around empty space) and therefore contributes nothing to $\beta_1$.
}
\label{fig:simplices}
\end{figure}

At the finest level, TDA assembles data points into simplices. A 0-simplex is a single vertex (\emph{e.g.} a grain centroid), a 1-simplex is an edge connecting two vertices, and a 2-simplex is a filled triangle among three vertices. Chains of simplices of the same dimension can be concatenated to form more complex objects. Figure~\ref{fig:simplices} shows examples of simplices of varying dimension alongside a 1-chain, a sequence of edges $\{ v_0,v_1 \}+\{v_1,v_2\}+\{v_2,v_0\}$ that forms a closed loop. Crucially, this particular 1-chain is also the boundary of the 2-simplex $\{v_0, v_1, v_2\}$, \emph{i.e.} the triangle fills it in completely. Because the loop bounds a higher-dimensional face, it does not encircle a genuine hole and thus contributes nothing to $\beta_1$. Consequently, in an effort to
quantify topological features, TDA is concerned with determining if cycles connect around loops in
the data as opposed to being boundaries of higher dimension simplices.

\begin{figure*}
    \centering
    \includegraphics[width=0.85\textwidth]{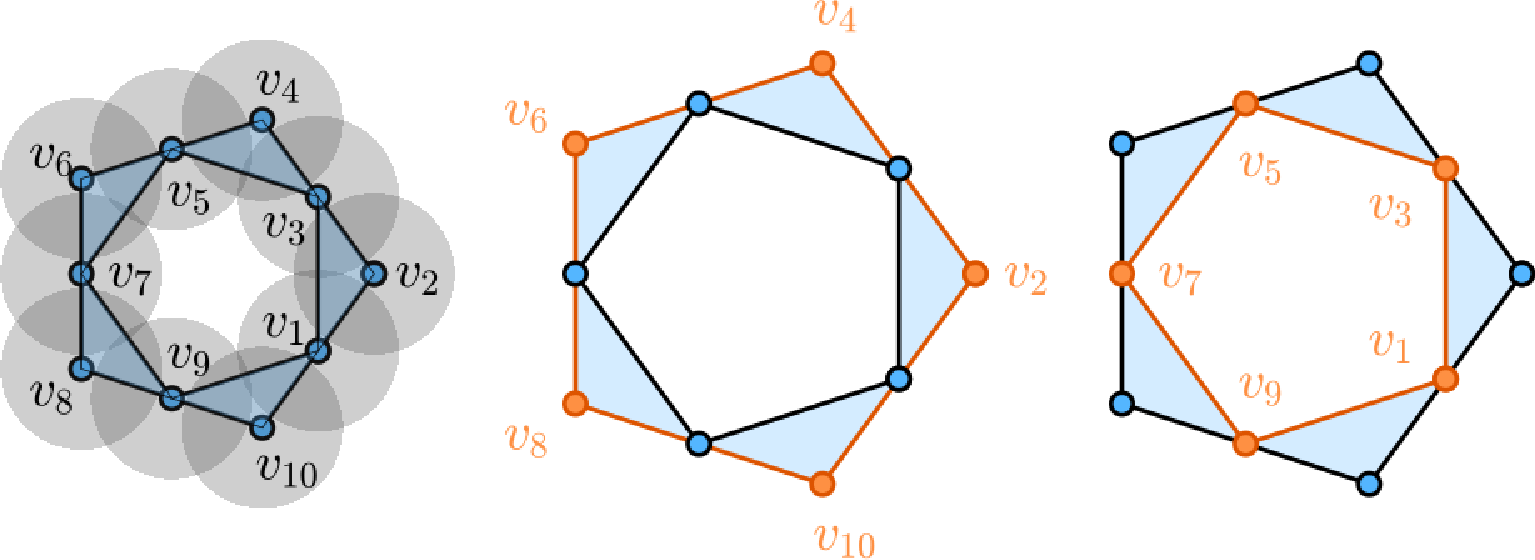}
\caption{For a given $\epsilon$-ball distance, (left) the vertices with even indices (middle) form an equivalent cycle to those with odd indices (right). Both cycles are example representations from a single equivalence class constituting the first homology group. That is, both highlighted $1$-cycles are equivalent up to boundary of the $2$-chain formed by five $2$-simplices that `loop' around a single, common hole in the data, \emph{i.e.} $\beta_1 = 1$. 
}
\label{fig:hole}
\end{figure*}

Figure~\ref{fig:hole} makes this distinction concrete at the scale of a small microstructure. From a set of 10 vertices at a given $\epsilon$, five 2-simplices can be formed along the outer boundary of the configuration, but the five 1-simplices forming the central ring cannot be filled in, \emph{i.e.} no 2-simplex spans the interior gap. This central ring is therefore a genuine 1-cycle that is not a boundary, yielding $\beta_1 = 1$. Moreover, the figure also illustrates that the cycle representative is not unique, as the two highlighted paths around the hole are homologous (they differ only by the boundary of the outer 2-chain) and TDA correctly identifies them as equivalent descriptions of the same topological feature.

In the context of grain microstructures, a ring of fine grains surrounding a coarse
central grain constitutes exactly such a cycle. A microstructure exhibiting necklacing
will therefore produce elevated $\beta_1$, making Betti-1 numbers the natural
topological descriptor for the spatial patterns described in~\cite{ASTM_E1181}. We focus
primarily on $\beta_1$ in the numerical experiments that follow, though $\beta_0$ is also informative in certain regimes. The formal algebraic construction of Betti numbers via chain groups, boundary operators, and homology quotient groups is provided
in Appendix~\ref{app:TDA_math} for the interested reader.

A single snapshot of Betti numbers at one fixed scale can be misleading --- a loop
apparent at a coarse resolution may simply reflect noise at a finer one. Persistent homology resolves this by tracking how Betti numbers evolve as a
continuous scale parameter $\epsilon$ varies. As $\epsilon$ increases, topological
features are born (a new loop forms) and eventually die (the loop is filled in by
denser connectivity). Each feature's lifetime is recorded as an interval $[\epsilon_{\mathrm{birth}},\, \epsilon_{\mathrm{death}}]$, and the collection of all such intervals is called a \emph{persistence barcode}~\cite{ghrist2008barcodes}. Features represented by long intervals (those born early and dying late) persist robustly across scales and are considered structurally significant, while short-lived features are treated as noise or fine-scale fluctuations. This multi-scale perspective yields an immediately interpretable summary of topological texture that is free of arbitrary threshold choices. The formal definition of homology groups and their induced chain maps is provided in Appendix~\ref{app:TDA_math}.

\begin{figure}
    \centering
    \includegraphics[width=0.45\linewidth]{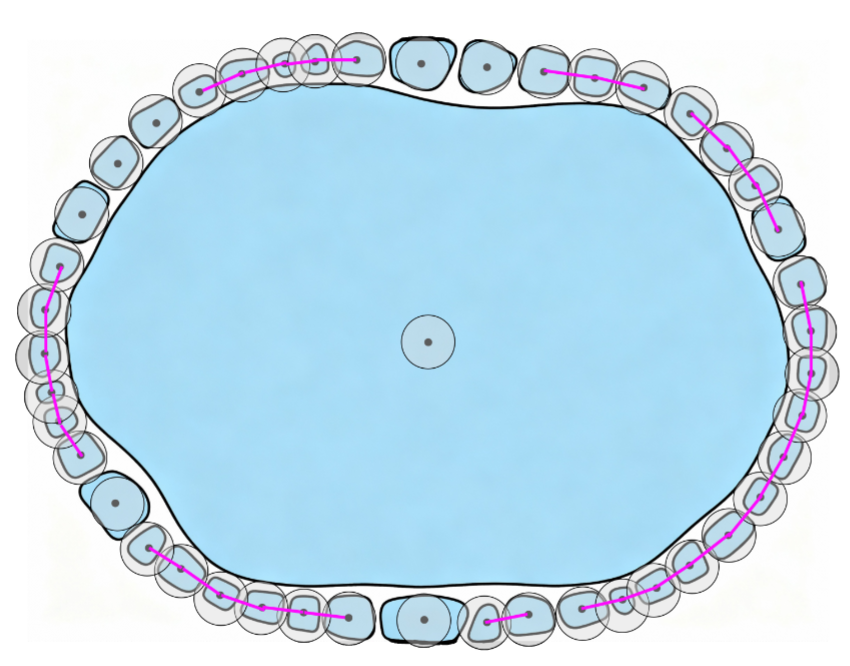}
    \includegraphics[width=0.45\linewidth]{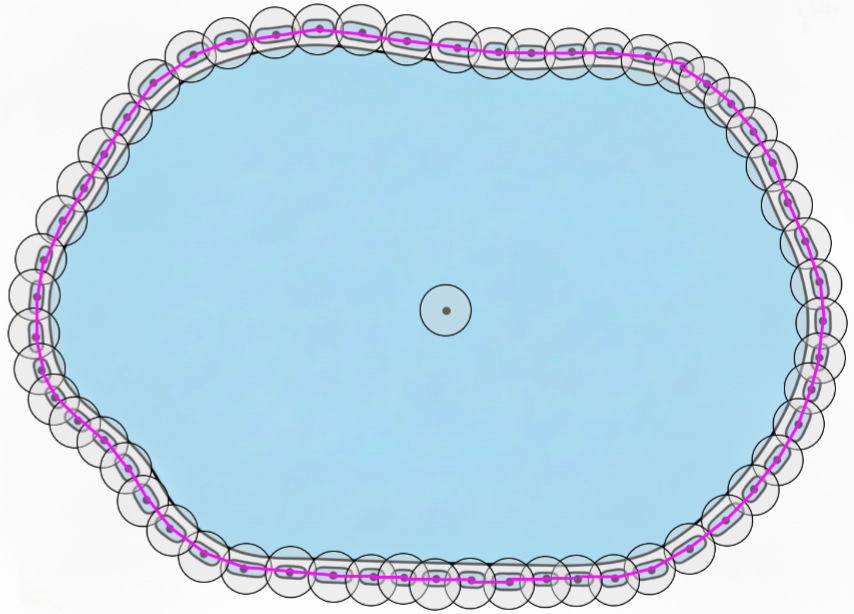}
    \caption{Illustration of necklacing and persistent holes using Rips complexes on grain centroids. In both panels, $\epsilon$-balls of equal radius are centered at grain centroids. Note the slight difference in grain size in the small grains between the two panels. In the high-necklacing case (right), many small grains form a finely graded ring around a larger grain, allowing the $\epsilon$-balls to connect and complete a cycle, producing a one-dimensional hole that is born relatively early and persists over a long range. In the low-necklacing case (left), fewer, larger grains form a coarser ring, so at the same $\epsilon$ the full cycle cannot be completed, delaying the birth and shortening the lifetime of the corresponding hole. In both configurations, the hole eventually dies at a similar $\epsilon$ value, when the $\epsilon$-ball of the large central grain intersects with $\epsilon$-balls of the surrounding grains.}
    \label{fig:rings}
\end{figure}

To apply TDA to a discrete set of grain centroids, we use the Rips-Vietoris filtration (Rips complex), a computationally efficient~\cite{zomorodian2010fast} filtration which quantifies spatial connectivity across scales. At a given scale $\epsilon$, an edge is drawn between two points within distance $\epsilon$ of each other, a triangle is filled whenever three pairwise distances are within $\epsilon$. As $\epsilon$ grows, more connections are added until the entire point cloud merges into a single component. In the context of grain structures, a Rips complex constructed on grain centroids has the capacity to detect necklacing patterns through the emergence and persistence of one-dimensional holes formed as a result of ring-like patterns of grains. A high degree of necklacing leads to such holes appearing early and persisting over a wider range of $\epsilon$, since cycles can be completed with relatively short connections between neighboring centroids. Conversely, when necklacing is weak or absent, the corresponding cycles can only form at larger $\epsilon$, if at all, because a larger neighborhood radius is required to connect the centroids all the way around the larger grain, resulting in shorter-lived or missing persistent holes. See Figure~\ref{fig:rings}.

Most classical TDA applications involve a single filtration variable, thus summarizing spatial connectivity alone. However, in the context of material microstructures, the richness of image patterns emerges not only from spatial arrangement but also from the intrinsic geometry of grain shapes. A key methodological question therefore arises: which filtration parameters best serve the quantification of pattern, and how can topological tools flexibly incorporate new metrics such as shape similarity? This motivates the dual-parameter framework described in Section 3, and requires first establishing a meaningful notion of shape distance.

\subsection{Shape Distance}

By focusing on the individual geometric properties of planar curves, shape analysis quantifies key characteristics, such as boundary ``roughness,'' and generalized scale, which influence a material's physical properties. 
Shape-based metrics are typically defined to capture invariances to rigid transformations, making them well-suited for precise measurement of highly varied and complex morphologies found in EBSD data.

A general framework for shape analysis represents each curve in an ensemble of planar curves as a collection of landmark points $n$, encoded in an $n\times 2$ matrix, and maps these collections onto matrix manifolds - \emph{e.g.} the Stiefel or Grassmannian. On these manifolds, distances between shapes are defined via the Riemannian geodesic distance induced by the chosen metric. In practice, direct computation of geodesics on a highly curved, high-dimensional manifold can be expensive, so we work in normal coordinates centered at an archetypal reference or template shape, typically the Fréchet mean of the ensemble. Within this coordinate system, the distance between a given shape and the mean shape is computed as the norm of the logarithmic map from the shape to the tangent space at the origin. Specifically, if $p$ is the reference point, and $z$ a shape on the manifold, the shape distance is given by $d_{\mathcal{M}}(p,z) = ||\text{Log}_p(z)||_g$ over a geodesic ball. Here the log map sends a shape $z$ to a tangent vector $\text{Log}_p(z)\in T_p\mathcal{M}$ at the reference $p$, and the norm is induced by the Riemannian metric $g$, reducing to the standard two-norm in these local coordinates where $p$ becomes the central shape at the origin.

This general framework allows for great flexibility: the specific choice of manifold and invariance structure depends on the application's needs, such as what transformations should be ignored or preserved. The same framework supports alternative metrics, such as elastic or information-geometric metrics, without altering the essential geometric ideas. Once a meaningful shape distance is established, it can serve as the foundation for a range of downstream analyses, including TDA.

In this work we use Separable Shape Tensors (SSTs)~\cite{grey2024explainable} as a computationally efficient realization of this framework. SSTs promote invariance to rigid body transformations, representing each grain boundary curve as a separable product $\widetilde X(t,\ell) = X(t) P(\ell)$ of two geometrically interpretable factors: $X(t)\in \mathbb{R}^{n\times d}_* / GL(d) \cong Gr(d,n)$ encodes undulation (the nonlinear, oscillatory character of the boundary curve) and $P(\ell)\in GL^+(d)/SO(d) \cong S^d_{++}$ encodes generalized scale (the anisotropic size and extent of the grain). The separability of these two components means that scale and undulation can be examined independently without conflating distinct geometric phenomena.

Leveraging fast implementations of rank-$d$ weighted-SVD facilitates rapid computations over an \textit{ensemble} of random matrices $\{ \widetilde{X}_k \}$ where $k$ potentially enumerates thousands or tens of thousands of shapes. With SST approximations $\{ \widetilde{X}_k \} \mapsto \lbrace ([\widetilde{X}], P)_k\rbrace$, we compute a mapping to normal coordinates of a central tangent space at the (intrinsic) Fr\'echet mean $([\widetilde{V}_0],P_0)$ over the corresponding product manifold $Gr(d,q)\times S^d_{++}$. The procedure, offered in detail in~\cite{grey2023separable}, executes in seconds on a conventional modern laptop. This computational efficiency makes the method well-suited to high-throughput microstructure screening pipelines, where hundreds of EBSD scans (such as those generated during combinatorial alloy development or in-situ process monitoring) must be characterized rapidly without the training overhead associated with deep generative models such as variational autoencoders.

Building submanifolds from parent matrix manifolds in this way means first representing each curve as a separable shape tensor $([\widetilde{X}], P)$ and mapping these tensors to normal coordinates at the Fréchet mean on the product manifold $Gr(d,q)\times S^d_{++}$. A low-dimensional, data-driven submanifold is then obtained by restricting to the linear subspace spanned by the dominant principal directions, which captures the main modes of shape variation while preserving the underlying Riemannian geometry. Since the geometry of the underlying manifolds is largely understood and supported by robust theoretical foundations, each computation involved in defining parameters of the submanifold can be explained using the interpretations of linear algebra and Riemannian geometry as opposed to ambiguous latent spaces of competitive generative models. Consequently, we inherit the following product metric as a distance between an arbitrary separable shape tensor, $([\widetilde{X}], P)$, and the intrinsic mean, $([\widetilde{V}_0],P_0)$,
\begin{equation} \label{eq:distance}
    \gamma([\widetilde{X}], P) \coloneqq d(([\widetilde{V}_0],P_0),([\widetilde{X}],P);a,b)= \sqrt{a \Vert \text{Log}_{[\widetilde{V}_0]}([\widetilde{X}])\Vert^2_F + b\Vert \text{Log}_{P_0}(P)\Vert^2_{P_0}}
\end{equation}
such that $\gamma:Gr(d,q)\times S^d_{++} \rightarrow \mathbb{R}_{\geq 0}$ represents SST distances from the fixed intrinsic mean, $([\widetilde{V}_0],P_0)$. Here, we take the Frobenius norm, $\Vert \cdot \Vert_F$, ~\cite{absil2008optimization} and norm induced by the affine-invariant metric, $\Vert_\cdot \Vert_{P_0}$ at $P_0$~\cite{fletcher2003statistics} for geodesic completeness. Additional computational details are available in~\cite{grey2023separable}. Positive weights $a$ and $b$ are selected empirically to normalize distances for the purposes of combining their distinct value ranges.

\begin{figure*}
    \centering
    \includegraphics[width=\textwidth]{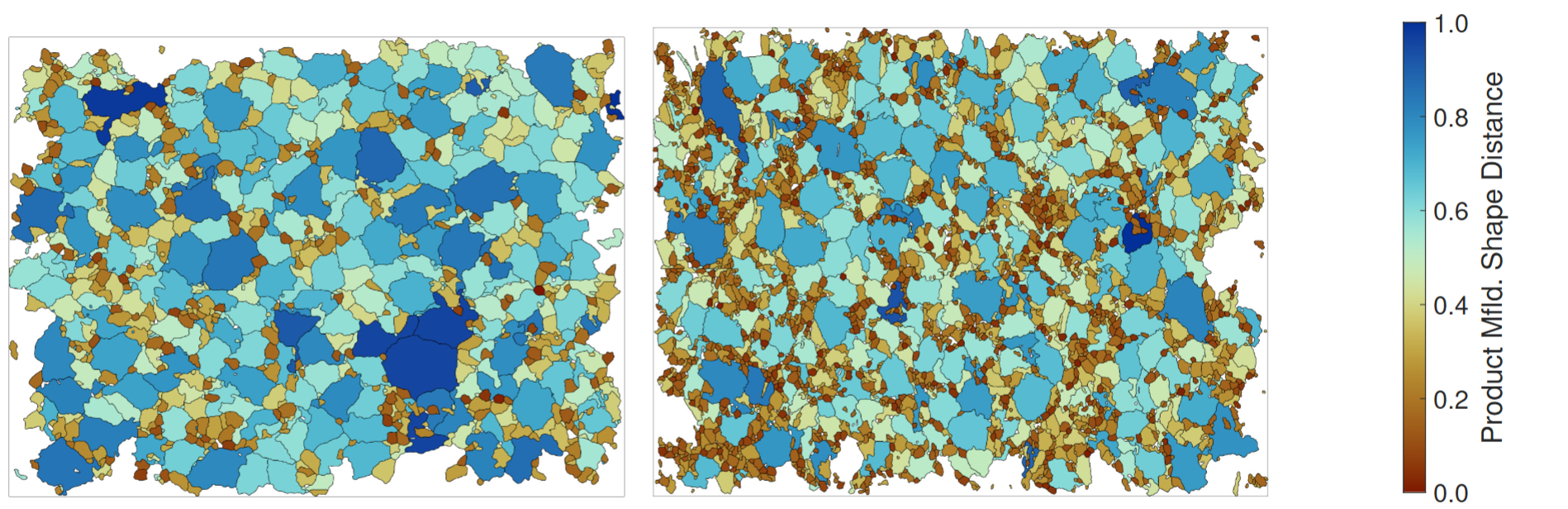}
    
    \caption{Two example microstructures of ice samples PIL184 (left) and  PIL185 (right) from ~\cite{Fan2020}. Grains are colored according to the (normalized) product submanifold shape distance of reparametrized grain boundary landmarks from the approximated origin, $([\widetilde{V}_0],P_0)$.
    }
    \label{fig:shape_dist}
\end{figure*}

Figure~\ref{fig:shape_dist} depicts a recoloring of the two microstructures in Figure~\ref{fig:eg_grains} according to~\eqref{eq:distance}. The product shape distance combines scale and undulation, so small (near 0) values of the color-map imply generalized anisotropic scales and undulations more similar to the average value, while larger values (near 1) indicate discrepancies from the average. The diverging color scheme reveals clear variation in the shapes distances in both example microstructures and an associated variation in spatial arrangement---\emph{i.e.}, smaller grains which appear to form connected components and loops around regions with larger grains in Figure~\ref{fig:shape_dist} (right). \textit{Our interest is now focused towards quantifying this obvious visual discrepancy in the spatial pattern of the images, left being very distinct from right}.

\begin{figure}[h!]
    \centering
    \includegraphics[width=\textwidth]{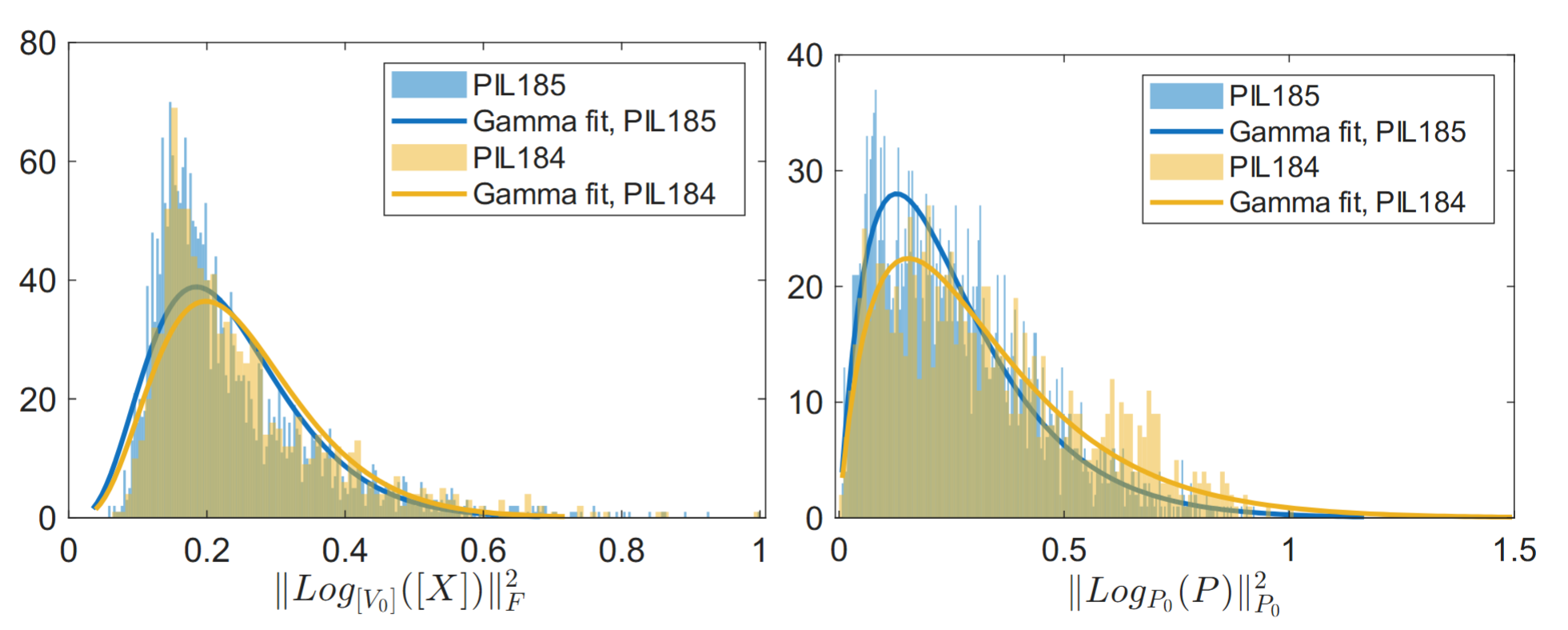}

    \caption{Histograms with fit distributions to visualize the distinction in shape undulation distance (left) and scale distance (right) metric distances between the two images being compared. Evidently, there is very little discrepancy in undulation distances while there are clearly discrepancies in shape scales. In particular, PIL184 data has more larger grain shapes.
    }
    \label{fig:dist_hists}
\end{figure}

Figure~\ref{fig:dist_hists} shows a histogram of the shape factors for PIL185 and PIL184.  The most significant differences are in the scale distance. While the histograms and fitted functions are discernibly different, and  the fitted distribution of $SPD$ distances are sufficient to achieve binary classification between these two microstructures~\cite{grey2024explainable}, extracted mean values would be harder to establish as statistically significant. However, this shape statistic completely ignores how shapes are spatially arranged within the image, an important property for materials characterization. Quantifying the obvious visual discrepancy in the distribution of shapes as well as their spatial patterns is precisely the aim of the bi-filtration framework introduced next.

\section{Combining TDA and SST}\label{sec4}

While single-parameter persistent homology conditioned only on spatial proximity of grain centroids can reveal nontrivial spatial organization through the persistence of one-dimensional holes, we emphasize that building simplices from grain centroids in this way has \textit{an obvious dependence on both the spatial arrangement and the geometric complexity (\emph{e.g.} scale and nonlinearity) of grain shapes}. Indeed, Figure~\ref{fig:grains_sublevels} illustrates this effect by showing Betti-1 persistence for three different shape-distance sublevel sets of PIL184. Depending on which subset of shapes is selected, the birth and death times of cycles can shift substantially. Thus, comparing and contrasting topological features of EBSD images will \textit{require a treatment of both the spatial arrangement informed by the Rips-Vietoris complex and the distance of shapes from an approximated intrinsic mean, $([\widetilde{V}_0],P_0)$}.

Shape descriptors like those of SST are especially well-suited for this purpose because they are invariant to rigid transformations—encoding purely intrinsic information about morphology, independent of position and orientation of curves in space. As a result, SST distances reflect only the intrinsic geometry of grains and are naturally agnostic to their global spatial arrangement. This makes them the ideal complement to spatial filtration parameters based on Euclidean distances between centroids. Their mutual independence ensures that a dual-parameter approach can meaningfully disentangle the impact of spatial arrangement from the underlying distribution of shape variations.

Just as examining a single value of $\epsilon$ in a Rips complex obscures the full evolution of topological features, focusing on a single threshold in shape distance misses how persistence varies across shape similarity. These observations motivate a framework that explicitly combines complementary shape and topological information, rather than treating them in isolation.

\subsection{Bifiltration Construction}
\label{subsec:bifiltration}

Given some features of interest for data there are potentially many ways to introduce a second parameter and extend a one parameter filtration to a two parameter filtration. However, this choice of a second parameter is highly dependent on the computational constraints and specific requirements of the problem. One very natural extension of one parameter persistence begins with the use of what are known as sublevel and superlevel filtrations.

The definition of a sublevel filtration is the following ~\cite{botnan2022introduction}: Suppose $\mathcal{M}$ is a topological space and $\gamma:\mathcal{M}\to \mathbb{R}$. The sublevel filtration $S^{\uparrow}(\gamma)$ is the $\mathbb{R}$ indexed filtration given by $S^{\uparrow}(\gamma) = \{p\in \mathcal{M}: \gamma(p)\leq \ell\}$. 
This definition allows us to define how to control another parameter within a bifiltration. Let \textbf{Simp} be the category of simplicial complexes and suppose that $\mathcal{P}\subseteq \mathcal{M}$ is a finite metric space such that $\gamma:\mathcal{P}\to \mathbb{R}$. The sublevel Rips bifiltration $S^{\uparrow}(\gamma): \mathbb{R}^2\to\textbf{Simp}$ is defined by $S^{\uparrow}_{\epsilon,\ell} = \gamma^{-1}(-\infty, \epsilon]$~\cite{botnan2022introduction}. 
The picture to keep in mind is a grid of simplicial complexes,
\begin{equation}
    \begin{matrix}
    &K_\epsilon^0 & \subseteq & \dots  & \subseteq & K_\epsilon^{\ell}\\
    &{\rotatebox[origin=c]{90}{$\subseteq$}}& & & &{\rotatebox[origin=c]{90}{$\subseteq$}}\\
    &\vdots & & \ddots & & \vdots\\
    &{\rotatebox[origin=c]{90}{$\subseteq$}} & & & & {\rotatebox[origin=c]{90}{$\subseteq$}}\\
    &K_{0}^0& \subseteq & \dots  & \subseteq & K_0^{\ell}
\end{matrix}
\label{eq:grid_complex}
\end{equation}
where the superscripts indicate level sets of $\ell$ and the subscripts are the Rips $\epsilon$-distance values. Each column is the usual Rips complex constructed on different sub(super)-level sets of data. 

The two axes of the resulting visualization have direct physical interpretations. The horizontal axis (shape distance $\ell$) tracks which grains are included in the analysis. At small $\ell$, only grains most similar to the mean grain are present; as $\ell$ increases, progressively more dissimilar grains (those that are larger, more undulating, or both)  enter the sublevel set. An example with three subensembles is shown in Figure~\ref{fig:grains_sublevels}. The vertical axis (spatial scale $\epsilon$) tracks the neighborhood radius at which those grains form connected rings. Small $\epsilon$ captures only nearest-neighbor connectivity, while large $\epsilon$ allows connections across longer distances. The Betti-1 count at each point $(\epsilon, \ell)$ in the grid therefore answers: ``Among grains whose shape is within distance $\ell$ of the mean, how many persistent spatial loops exist at scale $\epsilon$?''

We remark that this approach differs from the usual bi-filtration setting that gauges persistence in both parameters simultaneously. Since we are working in a single fixed normal coordinate frame over a Riemannian manifold of shapes, the only meaningful notion of distance is the radial distance from some intrinsic mean. Crucially, we are not searching for topological persistence in the shape-space distribution of planar curves. Instead, we are searching for topological persistence in the spatial arrangement of geometrically `similar' shapes, and additionally examining how that persistence changes as shapes depart from a common intrinsic mean. Nevertheless, by examining how persistence changes across these complementary filtrations, we can effectively capture a description of texture that factors both intrinsic fluctuations in shapes and their spatial arrangement.

\subsection{Interpreting the Bi-filtration}
\label{sec:Interpret}

The bi-filtration construction in~\ref{subsec:bifiltration} may be naturally understood as a surface, where the Betti-1 count $\beta_1(\epsilon,\ell)$ at each grid point defines a height which can be rendered as a filled contour map (see Figure~\ref{fig:mesa_examples}). These visualizations are essentially extensions of one parameter barcodes to bounded rectangular regions, allowing one to measure `bipersistence' of topological features over two independent parameters. The topography of the surface immediately reveals elevated areas of $\beta_1$ which identify where topological structure is concentrated.

In general, one expects a characteristic layout across the surface. Near the origin, where only the shapes closest to the mean are admitted and the spatial scale is small, connectivity is too sparse to complete cycles, so $\beta_1$ remains near zero. Moving toward larger $\ell$ at fixed small $\epsilon$, admitting more dissimilar shapes typically makes cycles easier to complete, producing a rise in $\beta_1$ even at fine spatial scales. At intermediate $\epsilon$, partial rings begin to close around larger neighboring structures, giving rise to modest, often localized elevations. $\beta_1$ reaches its highest and longest-lived values when enough shape diversity has entered the sublevel set and $\epsilon$ is large enough for rings to close, but not so large that they are filled in. These regions appear as a hotspots or extended plateaus in the surface. In practice, this layout allows a practitioner to locate, at a glance, the specific combination of shape distance and spatial scale at which a microstructure's topological texture is most pronounced.

\begin{figure}
    \centering

    \includegraphics[width=\textwidth]{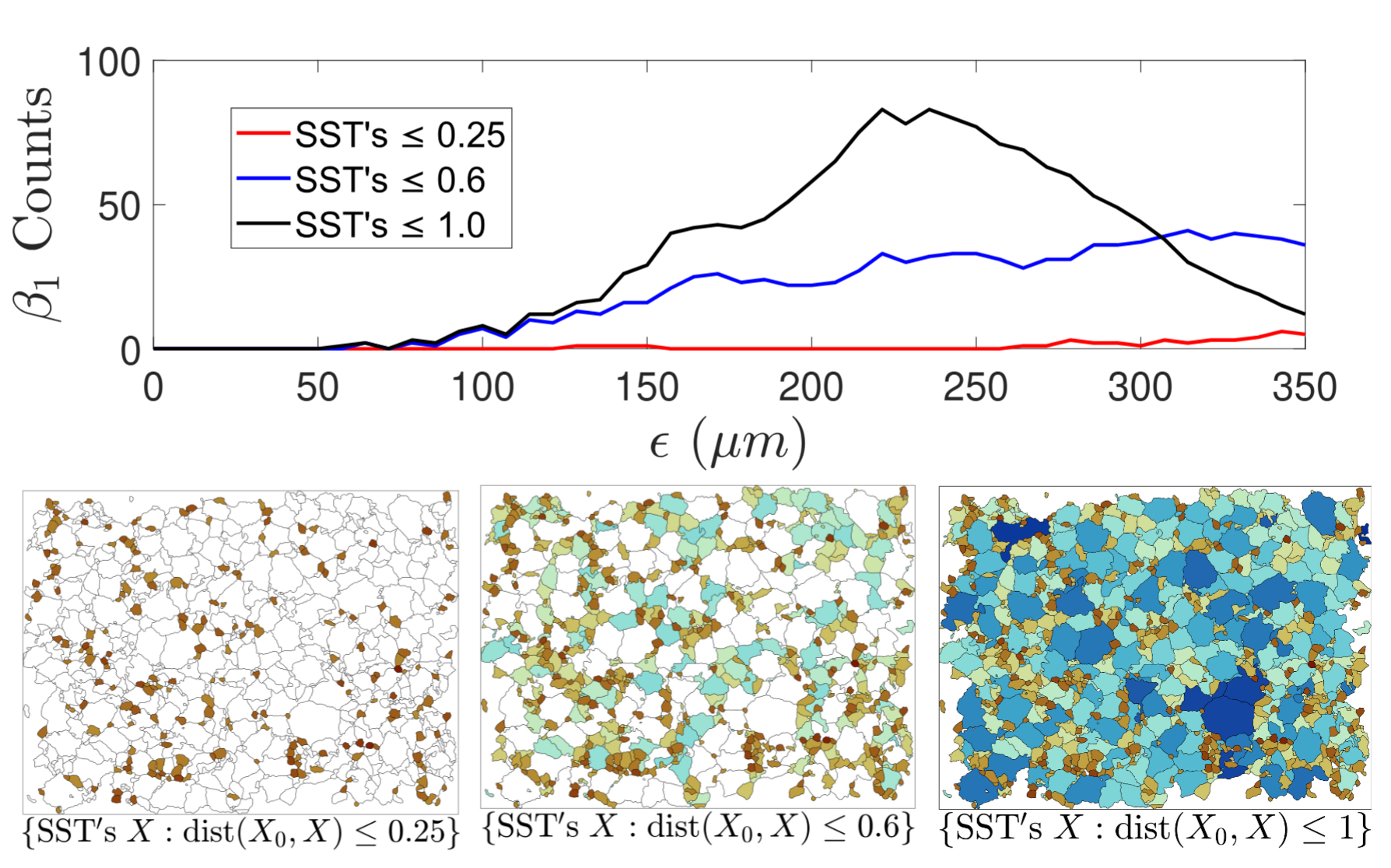}
    \vspace{3mm}
    \caption{Increasing sublevels of the product shape distance over the learned product manifold where the radius from the intrinsic mean determines how many shapes are included in each sublevel set. As the radius increases, additional, more distant shapes are incorporated into the analysis. Shapes furthest from the mean are characterized by being larger, more undulating, or a combination of both, capturing the diversity in scale and nonlinearity observed in the data.}
    \label{fig:grains_sublevels}
\end{figure}

To frame the results, a few examples of idealized microstructures are considered. If the microstructure contains homogeneous grain shapes that are randomly (uniformly) distributed, the bifiltration plot would have a single maxima at an x-axis value of 0.5 as the shape distance will not vary due to the homogeneity. The y-axis value $\epsilon$ of the maximum will be approximately equal to the mean grain radius.  If the grain shapes were perfectly uniform in shape and topology, the $\beta_1$ counts would be zero, as there is no persistence (cycles emerge but immediately become higher dimension simplicies and are excluded). For a second example, heterogeneous grain shapes that are randomly distributed are considered. In this case, the maxima will be elongated along shape distance and the $\beta_1$ counts will increase and then decrease as cycles are created throughout the range of grain shape distances and then are excluded as higher dimension simplicies form. This example is shown in Figure~\ref{fig:grains_sublevels} for SST's $\leq 1.0$, with a maxima around $\epsilon \approx 225\mu m$. For a bimodal distribution of grain shapes that form a topological necklace structure, such as shown in Figure~\ref{fig:rings}, once a cycle is completed, the $\beta_1$ counts will plateau and stay constant for a large range of $\epsilon$ values. This plateau is the topological signature of necklacing where a ring of fine grains forms a stable, persistent hole at intermediate spatial scales before it is eventually filled in as the $\epsilon$-ball of the large central grain intersects with those of the surrounding grains. An example is shown in Figure~\ref{fig:grains_sublevels} for SST's $\leq 0.6$, where a plateau starts at  $\epsilon \approx 250 \mu m$. 

Beyond visual comparison of the bi-filtration plots, one natural route towards a single scalar summary is to integrate the Betti-1 count over the bi-filtration domain. That is, to compute the area under the contour surface $\beta_1(\epsilon, \ell)$:
\begin{equation}
    \mathcal{I} = \int_0^{\ell_{\mathrm{max}}}\int_0^{\epsilon_{\mathrm\max}}\beta_1(\epsilon, \ell)d\epsilon d\ell.
    \label{eq:scalar_I}
\end{equation} This quantity $\mathcal{I}$ accumulates the total Betti-1 persistence across all shape-distance sublevel sets and spatial scales simultaneously, and can be interpreted as a measure of the overall topological complexity of the microstructure, weighting configurations by how many persistent loops they contain and over how large a region of the bi-filtration domain those loops are active. In practice, the integral is computed as a discrete sum over the grid of bi-filtration values, making it straightforward to evaluate from the same data used to produce the contour plots.

The scalar value $\mathcal{I}$ is sensitive to the \emph{extent} of necklacing and offers a natural basis for ordering or clustering EBSD samples. One may further refine this idea by computing the integral over a restricted subdomain of the bifiltration — for instance, integrating only over the low-to-moderate $\ell$ region where necklace-type islands are expected to concentrate. This yields a necklace-specific summary statistic that is tuned to the topological signature of necklacing and may offer improved sensitivity relative to the full integral when the primary goal is detecting or quantifying necklacing in the sense of ASTM E1181. More generally, one could define a family of such region-specific integrals corresponding to different topological patterns of interest (necklacing, banding, duplex distributions) each isolating a characteristic region of the $(\epsilon, \ell)$ domain.

\section{Results}\label{sec5}

\begin{figure*}[h!]
    \centering
    \includegraphics[width=\linewidth]{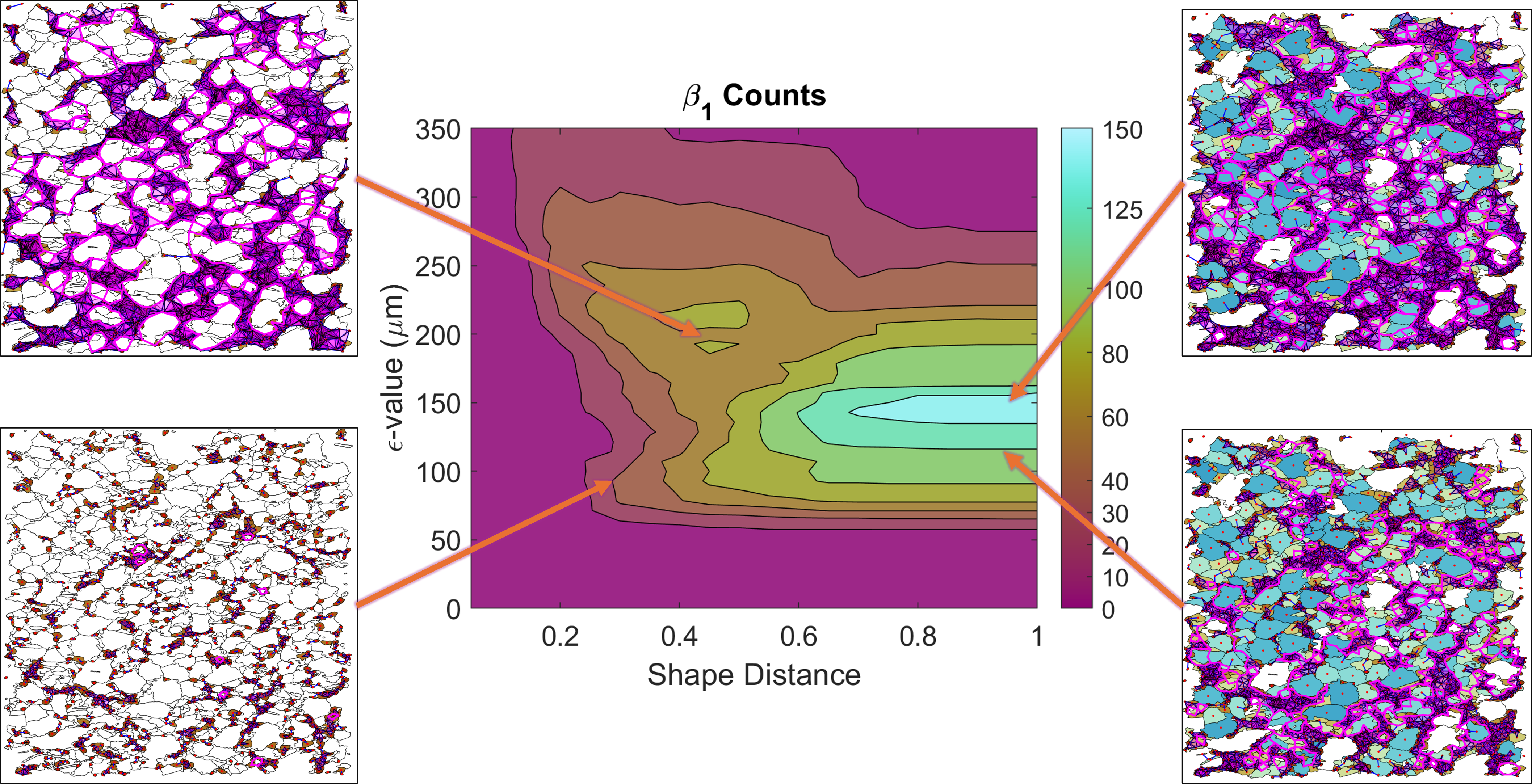}
    \caption{Example bifiltration of PIL 185 as filled contour map.  Four selected bifiltration points for PIL 185 are shown as insets to the figure.  Grain boundaries are shown as thin black lines, and grains with the shape distance equal or less than the value indicated are filled using the shape product colormap used in Figure~\ref{fig:shape_dist}. Rips complexes of representative 1-cycles (members of $H_1$) are highlighted in magenta. Equivalent cycles may be considered by deforming/contracting through filled magenta areas representing higher dimensional simplices, similar to Figure~\ref{fig:hole}. }
    \label{fig:mesa_examples}
\end{figure*}

Figure~\ref{fig:mesa_examples} is an example bi-filtration of PIL185 using the sublevel Rips construction of Section~\ref{subsec:bifiltration}, rendered as a filled contour map. As in~(\ref{eq:grid_complex}), the vertical axis represents $\epsilon$-values defining a Rips complex constructed over subsets of grain centroids, and the horizontal axis represents the shape-distance sublevel $\ell$ indexing which grains are included. Four selected points in the bifiltration are shown as inset panels confirming the expected behavior established in~\ref{sec:Interpret} on real data.

In the lower-left inset, few shapes and small spatial scales result in near-zero Betti-1 counts.  Moving horizontally to the lower-right panel, the number of Betti-1 holes towards the bottom of the plot increases as additional grains enter the sublevel set. The upper-left panel illustrates an intermediate $\epsilon$-band where rings of small grains begin to form chains encircling
larger grains, producing Betti-1 holes, but fewer in number than the lower-right panel. The
upper-right panel highlights a hotspot where the number of Betti-1 holes is maximal and their
lifetimes are longest, manifesting as a band of elevated Betti-1 intensity persisting over an extended vertical range of $\epsilon$. Moreover, the pattern of these hotspots exhibits nontrivial persistence across adjacent columns, indicating a clear dependence on shape distance in the horizontal direction.

\begin{figure*}[h!]
    \centering

    \includegraphics[width=\textwidth]{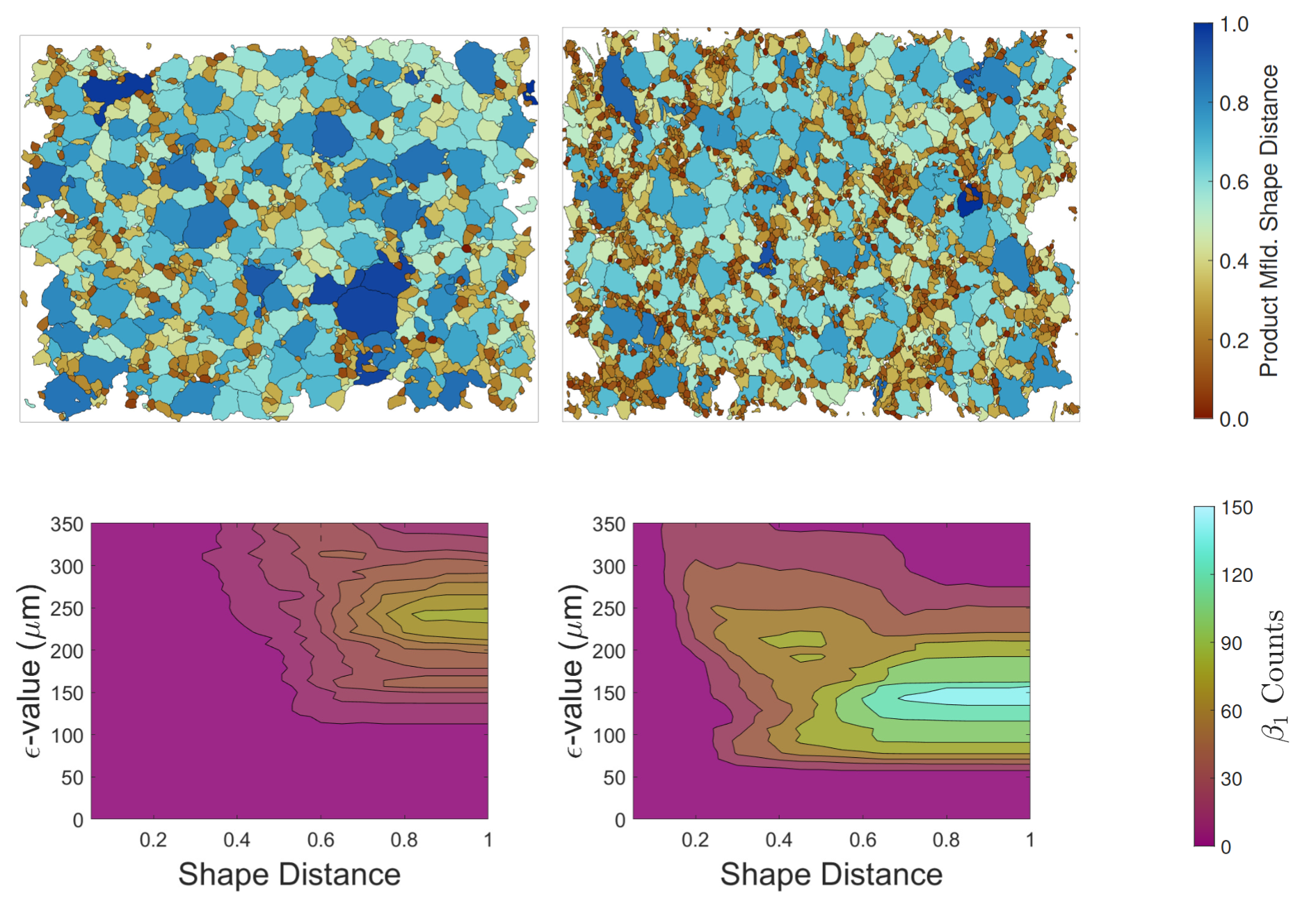}

    \vspace{3mm}
    \caption{Bifiltration filled contour image created from the example microstructures of ice samples PIL184 (left) and  PIL185 (right) from ~\cite{Fan2020}. Persistent homology is calculated using the open source software Javaplex~\cite{adams2014javaplex} with grain centroids as point cloud data. Visually it is clear that the contour profiles of Betti-1 counts for PIL 184 and PIL 185 exhibit apparent differences in both vertical and horizontal directions, indicating distinct topological features.}
    \label{fig:mesa_grains}
\end{figure*}

Figure~\ref{fig:mesa_grains} shows the bifiltrations 
constructed on the two scans of ice samples from above. Visually, it is readily apparent that these two samples demonstrate very different bipersistence structure. The bifiltration created from the PIL184 microstructure Figure \ref{fig:mesa_grains} (left), shows that the maximum number of persistent Betti-1 holes appear at an $\epsilon \approx 250 \mu m$ and shape distances from $0.6$ to $1$, while the PIL185 microstructure in Figure \ref{fig:mesa_grains} (right), has a much greater maxima over an interval of $\epsilon \approx 75 \mu m$ to $\epsilon \approx 175 \mu m$. The method not only establishes that PIL185 has a wider distribution in the bifiltration, with maxima in Betti counts occurring at lower shape distance values, but also confirms that the small grains create holes at intermediate radii rather than merely blending into the distribution. In Figure~\ref{fig:mesa_grains} (right), this effect appears as a localized ‘island’ of persistent holes at intermediate radii in the bipersistence plot, separated from the background by pronounced contour transitions. In contrast, when the small grains are distributed randomly or uniformly throughout the sample, these holes are filled in more uniformly, leading to a smoother, more homogeneous contour field without prominent topological features at any given scale. In other words, rather than being distributed randomly, the fine grains in PIL185 are organized in such a way that they form robust, ring-like patterns at certain spatial scales. This structured patterning directly corresponds to the heuristic notion of necklace topological features described in ASTM E1181. Thus, our approach provides for the first time a quantitative, objective evaluation of the type and degree of topological variation present. 

\subsection{ Strain Progression: Bi-filtration as Microstructure Fingerprint}

\begin{figure}
    \centering
    \includegraphics[width=\linewidth]{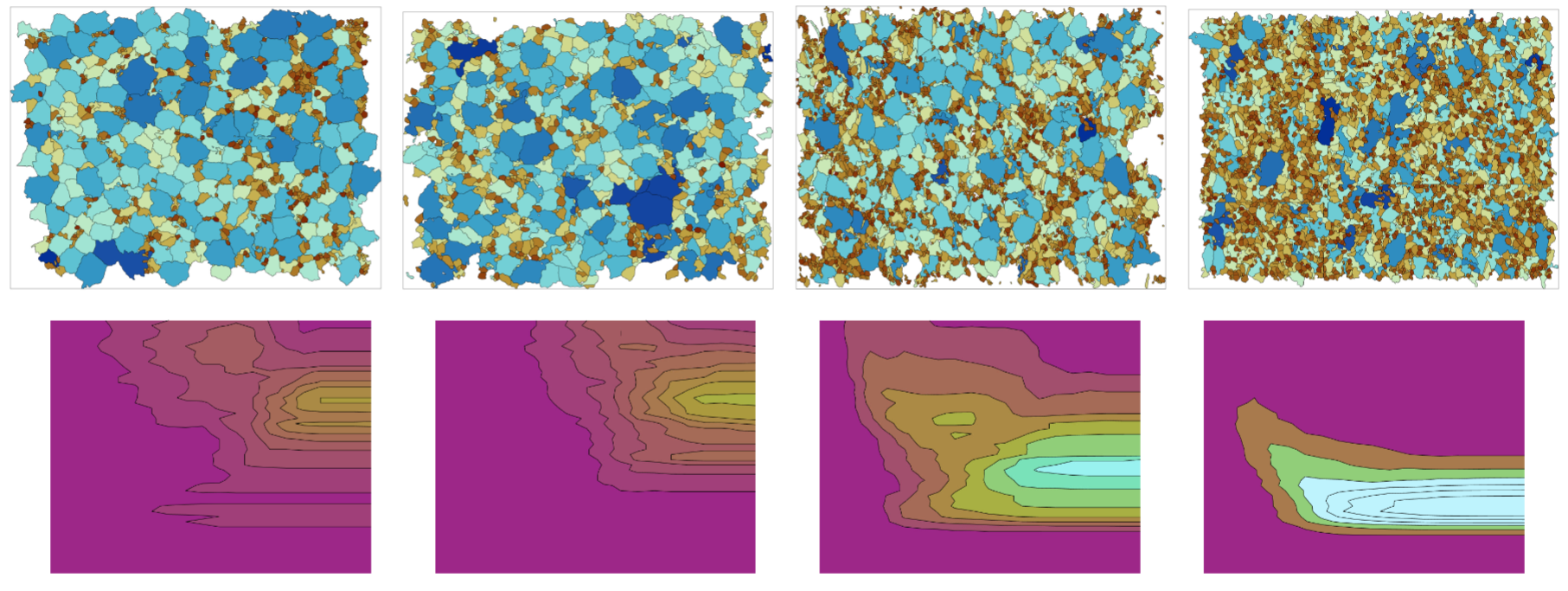}
    
    \caption{Ice microstructures with increasing strain are displayed from left to right as PIL254 (0.03 $\frac{mm}{mm}$), PIL184 (0.08 $\frac{mm}{mm}$), PIL185 (0.12 $\frac{mm}{mm}$), and PIL255 (0.20 $\frac{mm}{mm}$), alongside their corresponding bifiltration diagrams. The diagrams highlight that higher strain leads to more numerous and longer-lived persistent cycles, which appear earlier (both vertically and horizontally) in the filtration. This indicates that microstructural connectivity and complexity increase systematically with strain.}
    \label{fig:strain_degree}
\end{figure}

An important variable dictating the progressive development of ice microstructures is the degree of strain present in the formation process. Our bifiltration analysis is remarkably well-positioned to detect and quantify the role strain plays in the microstructure through persistent homology. Figure~\ref{fig:strain_degree} shows the bifiltrations for four ice microstructure scans arranged in order of increasing axial strain. As strain increases from the PIL254 (0.03 $\frac{mm}{mm}$), PIL184 (0.08 $\frac{mm}{mm}$), PIL185 (0.12 $\frac{mm}{mm}$), and PIL255 (0.20 $\frac{mm}{mm}$) datasets there is a  clear, progressive increase in both the number and `width' (lifetime) of persistent cycles in their corresponding diagrams. For low-strain samples, persistent cycles are fewer and short-lived, reflecting early-stage structures with large, equiaxed grains and minimal fragmentation. As strain progressively increases, cycles begin sooner and persist longer through the filtration, indicating the emergence of more complex features associated with dynamic recrystallization. The bifiltration diagrams indicate the PIL254 (0.03 $\frac{mm}{mm}$) and PIL184 (0.08 $\frac{mm}{mm}$) microstructures have little substantive change in shape distance or topology, as the diagrams are similar.  There is an abrupt change between the PIL184 (0.08 $\frac{mm}{mm}$) and PIL185 (0.12 $\frac{mm}{mm}$) diagrams, which demonstrates this is a critical strain range for recrystalization.

\begin{figure}
    \centering
    \includegraphics[width=0.5\linewidth]{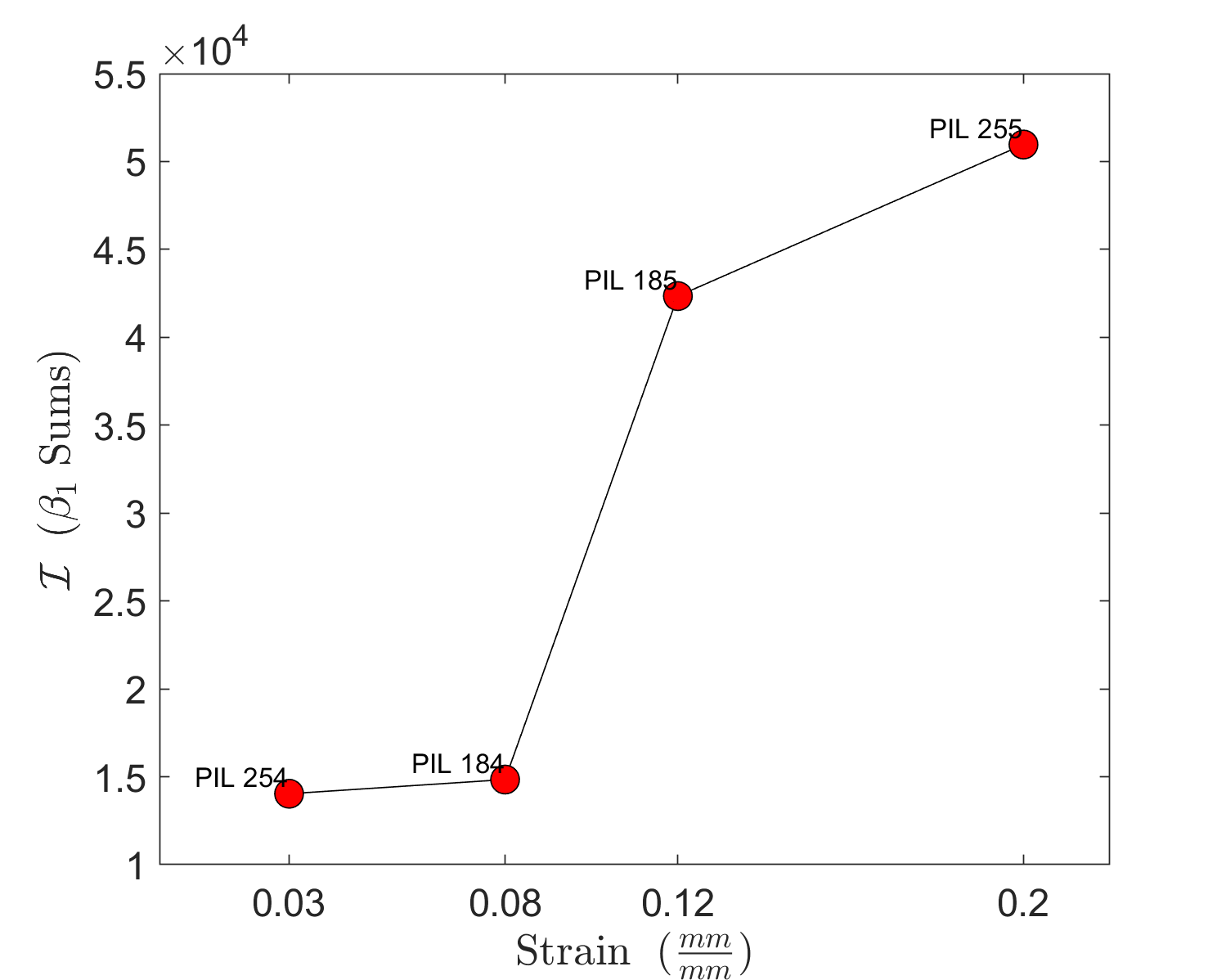}
    \caption{Scalar bipersistence summary $\mathcal{I}$ (equation~\ref{eq:scalar_I}) as a function of applied axial strain for the four ice microstructure samples PIL254 (0.03 $\frac{mm}{mm}$), PIL184 (0.08 $\frac{mm}{mm}$), PIL185 (0.12 $\frac{mm}{mm}$), and PIL255 (0.20 $\frac{mm}{mm}$). 
    }
    \label{fig:strain_vs_I}
\end{figure}

The scalar summary $\mathcal{I}$ from equation~\ref{eq:scalar_I} provides a direct quantitative confirmation of this trend across the four strain levels. Figure~\ref{fig:strain_vs_I} shows that $\mathcal{I}$ increases monotonically with strain in this range, consistent with the progressive intensification of necklace-type persistence observed visually in the corresponding bi-filtration surfaces. Between PIL254 and PIL184 (strain $3\%$ to $8\%$), $\mathcal{I}$ rises only modestly, indicating that the early stages of deformation/recrystallization produce relatively little change in the extent or persistence of Betti-1 structure. Between PIL184 and PIL185 (strain $8\%$ to $12\%$), however, $\mathcal{I}$ increases sharply, pointing to a comparatively abrupt transition in topological texture over this strain range. This mechanistic link between strain and persistent topological features in the bifiltration is consistent with prior findings by Fan et al., who specifically documented the evolution of ice microstructure as a function of increasing strain. In \cite{fan2020temperature}, they report that mean grain size is essentially unchanged over the $3\%-8\%$ strain range (114 $\mu m$ to 122 $\mu m$) and no core-and-mantle (necklace) structure is yet present. In the $8\%-12\%$ range, mean grain size collapses from 122 $\mu m$ to 75 $\mu m$, the number density of distinct grains more than doubles, and core-and-mantle microstructures first become apparent. The plateau behavior bewtween $12\%-20\%$ is consistent with the comparatively modest further change (75 to 64 $\mu m$) over the same interval, indicating a deceleration in grain refinement. In particular, \cite{fan2021using} analyzed grain-boundary sphericity on these same four samples and found that the estimated recrystallized area fraction continues to rise over this later interval ($33.5\%$ to $64.9\%$), even as the spatial arrangement of the larger, remnant-grain population stabilizes under what they attribute to a balance between grain-boundary migration and nucleation. This suggests the plateau in $\mathcal{I}$ reflects a saturation of the topological signature of necklacing rather than a halt in recrystallization itself.

Physically, the rising-then-plateauing behavior of $\mathcal{I}$ may reflect an initial phase of active grain-boundary reorganization and necklace diversification, followed by a transition in which further recrystallization is absorbed into an already-established necklace topology rather than generating new topologically distinct loops. Future work could confirm this by extracting persistent cycles and overlaying them on EBSD images to identify which grain rings drive the effect. Nevertheless, by mapping cycle counts, widths, and onset thresholds to strain magnitude, our methodology provides novel, quantifiable insight into the processes driving textural evolution and the mechanical behavior.

\section{Discussion}\label{sec6}
When compared with prevailing standards in material characterization, the current methodology offers distinct advantages in rigor and interpretability. In contrast to hand-picked or loosely quantified shape descriptors, the proposed topological approach provides a quantitative account of structural features that can be directly linked to the macroscopic properties of the material. Necklace-type spatial arrangements, quantified here through elevated and persistent Betti-1 counts, are mechanistically significant because fine-grain rings at coarse-grain boundaries are known sites of strain localization and grain-boundary sliding during creep and superplastic deformation~\cite{JULLIEN2024146927, MIKHAYLOVSKAYA2022142524}. Consequently, the scalar bipersistence summary $\mathcal{I}$ (\ref{eq:scalar_I}) offers a direct, quantitative proxy for microstructural features historically associated with reduced ductility and altered creep resistance in polycrystalline materials, providing a concrete bridge between topological texture and downstream mechanical performance. Subtle differences in spatial texture and associated Betti number distributions map closely to known variations such as necklacing and strain, supporting a numerically robust microstructure fingerprint. Furthermore, this analysis does not rely on any underlying assumptions about the form and distribution of grains nor any labeled data whatsoever, the presented observations are entirely intrinsic to the imaging.

The bifiltration approach developed here represents a tailored version of two-parameter persistent homology, adapted specifically to the constraints and opportunities presented by material microstructure data. Specifically, we leverage an axis of radial shape distance from an intrinsic mean rather than a pairwise shape distance. This property distinguishes our bi-filtration from classical settings in which both parameters may reflect spatial aspects; here, the decoupling of shape and spatial information enables the robust identification of persistent features that arise from their interaction or independence. While our construction may not necessarily reflect standard two-parameter TDA, it serves as a practical and illustrative example of how various filtrations, chosen to suit domain-specific questions, can be composed within the persistent topology framework.

For example, one can imagine constructing the bifiltration using solely the symmetric positive definite (SPD) scale term as shape distance (the second term in \ref{eq:distance}), omitting the first term in product distance that deals with non-linear deformations of curves. This perspective is particularly suitable for applications in which undulation or non-linear fluctuations of shape are either noise or not regarded as essential for the morphology under investigation. By focusing only on the SPD distance, the filtration becomes sensitive to intrinsic shape size and scaling variations, providing direct tools for contexts like grain growth or phase transformation where these geometric properties dominate.

Our results are predicated on certain constructions of shape and governing assumptions but serve as an effective demonstration for merging TDA and shape analysis. While these choices offer robust and interpretable frameworks for shape distance, it is crucial to emphasize that the mathematical machinery does not depend uniquely on any one definition of invariance or on a particular geometric setting. Alternate shape distances, such as those induced by elastic, information-geometric, or kernel-based metrics, may be substituted when required by the application or the characteristics of the data. Regardless of the metric chosen, the principles of mapping shapes into a suitable analysis space and quantifying their distance based on intrinsic geometric or functional structure remain broadly applicable, providing a foundation for many modern approaches in pattern recognition and statistical shape analysis.

This paradigm invites further research into the systematic testing of different permutations and combinations of filtrations (\emph{e.g.}, spatial proximity, orientation, alternative shape metrics) depending on the scientific context. As such, it provides a blueprint for adapting persistent homology analyses, enabling focused exploration of structural features relevant to various applications. With these observations and demonstrations in hand, scientists and engineers can use their creativity to bolster existing standards; creating the next generation of affordable, safe, and reliable materials based on more comprehensive assessments and correlations.

\section{Future Research}\label{sec7}

The integration of multi-parameter persistence into TDA represents a cutting-edge frontier in the field, but it remains relatively nascent---practical implementations leveraging two or more parameters are still emerging. Unlike single-parameter persistence, where invariants like Betti numbers and persistence diagrams provide clear and intuitive summaries, multi-parameter persistence requires more complex and nuanced representations. Exploring the use of three or more parameters could provide even deeper insights into the structure of high-dimensional data, but will require substantial theoretical and numerical innovations towards visualization and interpretation to balance the curse of dimensionality.

We look to future assessments involving high-dimensional features for extracting relevant information revealing new insights about patterns and properties of complex datasets that are not apparent through lower-dimensional analyses. We also remain curious about the use of non-uniform $\epsilon$-balls in a Rips complex. With smaller $\epsilon$ values in areas of high density, we can capture finer details in dense regions while preserving broader structures in sparser areas. Developing inter-level tri-filtrations~\cite{botnan2022introduction}, a direction-agnostic variant of sub(super)-level bifiltrations, could also enhance the analysis. This approach aims to capture intricate topological features without being constrained by a specific direction, potentially leading to more comprehensive descriptions of data.

Beyond the exploration of multi-parameter (\emph{e.g.}, bifiltration) persistent homology, it is equally compelling to consider how one might recast this richness back into a single parameter theory by shifting the underlying combinatorial structure. Specifically, rather than simplicial complexes where simplices are the core building blocks, one could instead adopt a polygonal complex where the fundamental units are the actual grain shapes treated as polygons. In this construction, the SST shape distance serves as the single filtration parameter, and connectivity is directly encoded by the grain boundaries: a 0-poly consists of an individual grain, a 1-poly is formed when two grains touch along their shared boundary, and a 2-poly naturally arises at 'triple points,' interpretable as convex hulls of shared interiors. This perspective not only generalizes the way features are assembled (potentially up to higher k-polytopes), but also creates an avenue by which the stratified connectivity of grain structures is encoded at the level of the physical microstructure. While the theoretical underpinnings of such polygonal complexes and their persistent (co)homology warrant further investigation, this approach elegantly ties the topology back to domain-specific features and suggests novel experimental directions for ongoing research and application.

Beyond descriptive characterization, the bi-filtration summary statistics developed here are well-suited as reduced-order input features for physics-based or data-driven simulation of microstructure evolution. For instance, $\mathcal{I}$ could serve as a calibration target or regularization term in phase-field or cellular automata models of grain growth and dynamic recrystallization, enabling simulations to be explicitly tuned to reproduce observed necklacing kinetics rather than relying on qualitative visual comparison alone. This positions the framework not only as an analysis tool but as a stepping stone toward topologically-informed simulation strategies for microstructure-sensitive design.

\ack{We greatly appreciate G\"unay Do\u{g}an at NIST for his guidance regarding applications and important data considerations as it relates to the nuances of processing EBSD imaging, and Adam Church at Colorado School of Mines for discussions around analysis of steel microstructures.  

The views expressed in the article do not necessarily represent the views of NIST or the U.S. Government. This work is U.S. Government work and not protected by U.S. copyright. The U.S. Government retains and the publisher, by accepting the article for publication, acknowledges that the U.S. Government retains a nonexclusive, paid-up, irrevocable, worldwide license to publish or reproduce the published form of this work, or allow others to do so, for U.S. Government purposes. This publication is intended to capture external perspectives related to NIST applied mathematics research. These external perspectives can come from industry, academia, government, and other organizations. This report was prepared as collaborative work; it is intended to document external perspectives.}

\funding{This work was partially supported by a National Institute of Standards \& Technology (NIST) Building the Future (BTF) internal funding initiative and the first author was supported in part by funds from the National Science Foundation (NSF) Graduate Research Fellowship Program (GRFP).}

\roles{The first and second authors contributed equally to this manuscript. The third author facilitated data interpretation, drafts, review, and edits.}

\data{Ice micrograph data is available at~\cite{Fan2020}.}


\bibliographystyle{unsrt}
\bibliography{bibliography}

\appendix

\section{TDA Formalism}\label{app:TDA_math}

\subsection{Homology}
Given a data set equipped with a notion of distance, a fundamental step in applying TDA is endowing it with a topological structure. A popular approach involves simplicial complexes as combinatorial structures that serve as discrete representations of topological spaces. Formally, a simplicial complex, $K$, is a finite collection of simplices, 
$
    \sigma_n = \{v_0, v_1, \dots, v_n\},
$
such that $\sigma_n \in K$ are collections of simple geometric objects enumerated by their vertices $v_0, v_1, \dots, v_n$.

Real-world data is embedded into these abstract structures by treating the points in a dataset as vertices ($0$-simplices) of a simplicial complex. This immediately raises the practical question: which points should be `connected' to form higher dimensional simplices? To address this, a parameter $\epsilon$ is introduced to control the scale at which topological features are detected. At a given $\epsilon$, we include connections between points whose separation is at most $\epsilon$, and exclude connections that exceed this threshold.

To make use of the group theoretic structure of homology, we construct chain groups on simplicial complexes. The $n^{th}$ chain group $C_n= \sum_m a_m\sigma_m$ is a free module with $n$-simplices as basis elements, and coefficients $a_m$ taken from the field $\mathbb{Z}_2$. 
Chains effectively allow us to `glue' together simplices as building blocks for more complicated geometric objects. 
Note that although the complex $K$ as a whole may contain simplices of various dimension, chain groups over $K$ may only be formed by considering simplices of the same dimension. Thus, depending on the dimensions of the simplices in $K$, many of the chain groups may be empty. 
See Figure \ref{fig:simplices} for examples of simplices of different dimension and a $1$-chain of $1$-simplices.

The boundary operator $\partial_n: C_n\to C_{n-1}$ is defined on $n$-simplices by taking the sum of their $n - 1$ dimensional `faces' with alternating sign, \emph{i.e.} 
$
    \partial(\{v_0, \dots, v_n\}) = \sum_{i=0}^n (-1)^i\{v_0, \dots, \widehat{v}_i,\dots v_n\},
$
extending linearly to the rest of $C_n$, \emph{e.g.}, $\partial_n(C_n) = \sum_m a_m\partial(\sigma_m)=: \text{Im}(\partial_n)$. The hat notation indicates that the $i$-th vertex is omitted. In this way the boundary map takes any $n$-chain group to a $n - 1$ chain group, inducing the sequence of maps enumerated by simplex dimension:
\begin{equation}
    \cdots \overset{\partial_{n+1}}{\longrightarrow} C_n \overset{\partial_{n}}{\longrightarrow} \cdots \overset{\partial_{2}}{\longrightarrow} C_1\overset{\partial_{1}}{\longrightarrow} C_0\overset{\partial_{0}}{\longrightarrow} 0.
\end{equation}

Chains in $\ker(\partial_n):=\{\sigma_m\in C_n:\partial_n(\sigma_m) = 0\}$ are called $n$-\textit{cycles} and represent complete ``loops'' formed by $n$-simplices. However, it is easy to show that the boundary of a boundary is empty and thus all boundaries are cycles\footnote{Formally we must utilize an ordering convention on the vertices.}, \emph{i.e.} $(\partial_{n}\circ \partial_{n+1})(\sigma) = 0$. Consequently, in an effort to quantify topological features, TDA is concerned with determining if cycles connect around loops in the data as opposed to being boundaries of higher dimension simplices. Thus, we define the quotient group $H_n:= \ker(\partial_n)/\text{Im}(\partial_{n+1})$, called the $n^{th}$ homology group by `dividing out' the boundaries of the simplices one dimension higher to identify cycles which are not boundaries. Figure~\ref{fig:simplices} illustrates a simple example showing how the 1-cycle is the boundary for a 2-simplex.

When $H_n$ is not empty, we can interpret individual elements as equivalence classes of cycles. The equivalence classes identify all $n$-chains with empty boundary (cycles) that are equivalent, or homologous, if they differ by a boundary in $\text{Im}(\partial_{n+1})$. This facilitates a formalism for collecting cycles that are also homotopically equivalent representations of a topological feature---\emph{i.e.}, sets of equivalent cycles which loop around a common hole in the data. Figure~\ref{fig:hole} illustrates and describes an example of a one dimensional $H_1$ homology group. From the set of 10 vertices and given $\epsilon$, five 2-simplexes can be formed at the boundaries, but a 1-cycle containing five 1-simplicies remains in the center. This set of verticies is akin to the `necklace' microstructure, with smaller (finer) cycles surrounding a larger (course) structure.  

The $n^{th}$ Betti number $\beta_n$ is the rank (number of generators) of the $n^{th}$ homology group and counts the $n$-dimensional 'voids' in the data. This characterization effectively allows identification of data spaces as having the same topology in the very broad sense of homotopy. 
Typically $\beta_0$ count distinct connected components formed by $0$-simplices and $\beta_1$ count distinct holes formed by $1$-simplices. In this application, \textit{spatial texture is represented by the number of equivalent cycles (Betti number) formed between grain centroids}. We focus on $\beta_1$ numbers in numerical examples but either may be of interest in applications.

\subsection{Persistent Homology}
To get an idea of which topological features are persistent, we introduce the concept of a filtration. Put simply, a filtration is a nested sequence of complexes induced by a parameter $\epsilon$ such that $K_i = K(\epsilon_i)$ and $\emptyset = K_0\subseteq K_1\subseteq\dots \subseteq K$ with $K$ the full complex. 

For $i<i'$ we have an inclusion map $K_i\hookrightarrow K_{i'}$ that induces a homomorphism on the respective chain groups
by composing each $n$-simplex in $K_i$ with the inclusion map to get a $n$-simplex in $K_{i'}$, then extending linearly to the rest of the chain group. This map sends cycles to cycles and boundaries to boundaries, and thus induces a homomorphism $f_n^{ii'} : H_n(K_i)\to H_n(K_{i'})$ on the respective homology groups~\cite{hatcher2002algebraic}.

The persistent homology groups are the images of $H_n(K_i)$ in $H_n(K_{i'})$, \emph{i.e.} the $n^{th}$ persistent homology group $H_n^{ii'} = \text{Im}(f_n^{ii'})$. The persistent Betti numbers are defined as the dimensions of these images, and extend the concept of Betti numbers to a multi-scale analysis. While Betti numbers tell us how many $n$-dimensional features exist in a topological space at a specific scale, persistent Betti numbers track how these features emerge as the parameter $\epsilon$ changes, giving us an immediate representation of how the homology of the complex is evolving. This is often visualized with what are called persistence diagrams, of which barcodes are common~\cite{ghrist2008barcodes}.

\end{document}